\def\be{\begin{equation}}
\def\ee{\end{equation}}
\def\bea{\begin{eqnarray}}
\def\eea{\end{eqnarray}}
\def\bse{\begin{subequations}}
\def\ese{\end{subequations}}

\documentclass[prl,twocolumn,showpacs,amsmath,amssymb]{revtex4}
\usepackage{graphicx}
\usepackage{dcolumn}
\usepackage{bm}
\begin{document}
\title{Breakdown of the perturbative renormalization group at certain quantum
       critical points
}
\author{D. Belitz$^1$, T.R. Kirkpatrick$^2$, and J. Rollb{\"u}hler$^1$}
\affiliation{$^1$Department of Physics and Materials Science Institute,
                  University of Oregon, Eugene, OR 97403\\
             $^2$Institute for Physical Science and Technology, and Department of
                 Physics\\
                 University of Maryland, College Park, MD 20742
         }
\date{\today}

\begin{abstract}
It is shown that the presence of multiple time scales at a quantum critical
point can lead to a breakdown of the loop expansion for critical exponents,
since coefficients in the expansion diverge. Consequently, results obtained
from finite-order perturbative renormalization-group treatments may be not be
an approximation in any sense to the true asymptotic critical behavior. This
problem manifests itself as a non-renormalizable field theory, or,
equivalently, as the presence of a dangerous irrelevant variable. The quantum
ferromagnetic transition in disordered metals provides an example.
%
\end{abstract}


\maketitle

The phenomenon known as critical slowing down is a dramatic effect near
critical points \cite{Hohenberg_Halperin_1977}. It occurs because the order
parameter (OP) relaxation rate $\tau$ diverges together with the correlation
length $\xi$. The dynamical critical exponent $z$ is defined by
$\tau\propto\xi^z$. If soft modes other than the OP fluctuations couple to the
latter, one needs to take them into account as well. For instance, at the
liquid-gas critical point one needs to consider shear modes
\cite{Hohenberg_Halperin_1977}, and the quantum ferromagnetic transition
\cite{Hertz_1976,Millis_1993} is influenced by soft particle-hole excitations
\cite{Kirkpatrick_Belitz_1996}. These additional soft modes are `generic' in
the sense that their softness is not related to the critical point (e.g., shear
modes in a fluid are soft due to momentum conservation), and they introduce
diverging time scales in addition to the critical one
\cite{Belitz_Kirkpatrick_Vojta_2004}. In contrast to the diverging length scale
$\xi$, which is unique in any given direction, one therefore usually has to
deal with multiple diverging time scales. In classical statistical mechanics
this complication remains restricted to the dynamics, as the static critical
behavior is independent of it, but in quantum statistical mechanics, where the
statics and the dynamics are coupled, the critical behavior of thermodynamic
quantities is also affected \cite{Hertz_1976}.

The technical description of phase transitions often relies on the
renormalization group (RG). There are two different formulations of the RG. The
first one, Wilson's momentum-shell RG \cite{Wilson_Kogut_1974, Ma_1976},
integrates out degrees of freedom with successively larger wave numbers in
order to derive a simpler effective theory valid only at asymptotically large
length scales, i.e., at criticality. The second one, often referred to as the
field-theoretic method, has its roots in high-energy physics. It introduces an
artificial `renormalized theory' in order to study the behavior of the physical
theory under changes of the momentum cutoff, i.e., the basic length scale
\cite{Zinn-Justin_1996}. These two formulations of the RG are believed to be
equivalent \cite{Weinberg_II_1996}. An important concept in the field-theoretic
formulation is the `renormalizability' of the theory, i.e., the possibility of
constructing a renormalized theory that is finite in the limit of an infinite
momentum cutoff by absorbing any infinities into a finite number of
renormalization constants. Renormalizability is crucial in order to ensure
scale independent RG flow equations and the power-law critical behavior that
generically follows from them. While this concept has no obvious analog in the
Wilsonian method, it is generally believed that lack of renormalizability can
be remedied by adding additional renormalization constants, which in the
Wilsonian method corresponds to the generation, under renormalization, of
additional terms that were not present in the original action. In either
method, one usually proceeds by performing a loop expansion in powers of a
suitable coupling constant. In conjunction with an $\epsilon$-expansion about
some critical dimensionality, this allows for a perturbatively controlled
determination of the critical behavior.

In this Letter we come to the remarkable conclusion that the perturbative RG
can break down at a quantum critical point if there are multiple divergent time
scales. Specifically, low-order results of an $\epsilon$-expansion for critical
exponents may be invalid because coefficients of terms that are formally of
higher order in $\epsilon$ diverge. A second conclusion is that
nonrenormalizability can be related to the existence of a dangerous irrelevant
variable (DIV) \cite{Fisher_1983} rather than to the generation of new terms in
the momentum-shell method. Our results are generic, but for the sake of
definiteness we illustrate them by using the quantum ferromagnetic transition
in disordered metals \cite{Hertz_1976} as an example. This transition is known
to have an upper critical dimension $d_{\text c}^+ = 4$
\cite{Belitz_et_al_2001a,Belitz_et_al_2001b}.
Accordingly, there is an effective quartic coupling constant
$u$ that is irrelevant with respect to a Gaussian mean-field fixed point (FP)
in spatial dimensions $d>4$, but becomes marginal in $d=4$. Standard lore
\cite{Wilson_Kogut_1974,Ma_1976} suggests that a controlled $\epsilon =4-d$
expansion of the critical exponents is possible for $d<4$. Indeed, we will show
below that to one-loop order one finds a flow equation
\be
\frac{du}{d\ell} = \epsilon u - \frac{11}{12}\,u^2 + O(u^3),
\label{eq:1}
\ee
where $\ell = \ln b$ with $b>0$ the RG length scale factor. This suggests a FP
value $u^* = 12\epsilon/11 + O(\epsilon^2)$, which determines the critical
exponent $\eta$ via $\eta = u^*/6 + O(\epsilon^2)$,
\bse
\label{eqs:2}
\be
\eta = 2\epsilon/11 + O(\epsilon^2).
\label{eq:2a}
\ee
The correlation length exponent $\nu$ and the dynamical exponent $z$ are given
in terms of $\eta$,
\be
z = 4 - \eta\quad,\quad\nu = 1/(2 - \eta).
\label{eq:2b}
\ee
\ese
These results have nothing in common with the exact 
critial behavior found in Ref.\ \onlinecite{Belitz_et_al_2001b}, which is not
even given by power laws. This creates an obvious conundrum, especially since
the $4-\epsilon$ expansion is a more standard and straightforward method than
the one employed in \cite{Belitz_et_al_2001b}.

Equations\ (\ref{eq:1},\ref{eqs:2}), although formally correct, do not
constitute an approximation to the true asymptotic critical behavior in any
sense. This can first be seen in the flow equations at two-loop order, where
the loop expansion parameter $u$ couples to an irrelevant variable $w$,
\bse
\label{eqs:3}
\bea
\frac{du}{d\ell} &=& \epsilon u - \frac{11}{12}\,u^2 - \frac{u^3}{32}\,(1-\ln
      w) + O(u^4),
\label{eq:3a}\\
\frac{dw}{d\ell} &=& -2w + \frac{11}{12}\,uw + \frac{u^2}{32}\,(1-\ln w)w +
O(u^3).\quad
\label{eq:3b}
\eea
\ese
$w$ is irrelevant with a scale dimension $-2 + O(\epsilon)$ and a FP value
$w^*=0$. It is common that some scaling function depends singularly on an
irrelevant variable, making the latter a DIV with respect to the corresponding
observable \cite{Fisher_1983}. Here, however, the coefficients of the loop
expansion itself depend singulary on $w$, leading to a breakdown of the
perturbative RG \cite{liquid_gas_footnote}. This phenomenon is reminiscient of
the breakdown of the virial expansion for transport coefficients in classical
statistical mechanics \cite{Dorfman_Kirkpatrick_Sengers_1994}. In either case,
a seemingly systematic expansion in powers of a coupling constant is ruined by
the appearance of logarithmic terms. Physically, $w$ reflects the ratio of the
diffusive time scale of the itinerant electrons, which is characterized by a
frequency scaling as a wave number squared, $\Omega\sim{\bm k}^2$, and the
critical time scale characterized by $\Omega\sim{\bm k}^4$
\cite{Hertz_1976,Hertz_footnote}.

In the remainder of this Letter we sketch the derivation these results from a
model field theory which has the same structure as the field theory that
describes the quantum phase transition in disordered itinerant Heisenberg
ferromagnets. In addition to demonstrating the origin of the breakdown of the
loop expansion manifest in Eqs.\ (\ref{eqs:3}), it will show that this
breakdown is technically related to the field theory being non-renormalizable
in a way that cannot be cured by introducing additional renormalization
constants. This non-renormalizability is equivalent to the dangerous
irrelevancy of $w$.

Let us consider a quantum field theory that couples diffusive excitations $q$
to a $\phi^4$-theory, with Gaussian propagators
\bse
\label{eqs:4}
\bea
\langle q_n({\bm p})\,q_{-n}(-{\bm p})\rangle &=& \frac{G}{{\bm p}^2 +
GH\vert\Omega_n\vert}\ ,
\label{eq:4a}\\
\langle \phi_n({\bm p})\,\phi_{-n}(-{\bm p})\rangle &=& \frac{1}{t + a{\bm p}^2
+ \frac{aC_1\vert\Omega_n\vert}{{\bm p}^2 + GH\vert\Omega_n\vert}}\ .
\label{eq:4b}
\eea
\ese
The fields $q$ and $\phi$ are functions of a wave vector ${\bm p}$ and a
Matsubara frequency $\Omega_n = 2\pi Tn$. The propagators in Eqs.\
(\ref{eq:4a}) and (\ref{eq:4b}) have structures appropriate for diffusive
particle-hole excitations and paramagnons, respectively, in disordered metals.
$t$ is the distance from a critical point for the $\phi$-excitation, and $G$,
$H$, $a$, and $C_1$ are model parameters. Notice that the propagators contain
the two time scales mentioned above. This will be crucial later. Motivated by
the known properties of the ferromagnetic problem \cite{Belitz_et_al_2001a} we
now assume that under renormalization, $t$, $G$, and $C_1$ are not singularly
affected, while $a$ and $H$ acquire momentum and frequency dependences,
respectively, that are determined by a set of coupled integral equations which
read, at quantum criticality, $T=t=0$, and with a momentum cutoff $\Lambda$,
\bse
\label{eqs:5}
\bea
a({\bm k}) &=& a + \frac{C_2}{3} \int_0^{\infty} d\omega\int_{\vert{\bm
k}\vert}^{\Lambda} dp\ \frac{p^{\,d-1}}{\left[p^2 + H(\omega)\omega\right]^3}\
.\quad
\label{eq:5a}\\
H(\Omega) &=& H + \frac{3C_2}{4}\int_0^{\Lambda} dp\ \frac{p^{\,d-1}}{a({\bm
p})p^{\,4} + a C_1\Omega}\ .
\label{eq:5b}
\eea
\ese
Here $C_2$ is another coupling constant.

\begin{figure}[t]
\includegraphics[width=7.5cm]{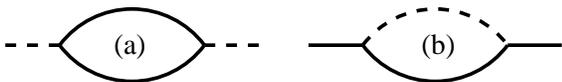}
\caption{\label{fig:1} Diagrammatic contributions to $a({\bm k})$ (a) and
$H(\Omega)$ (b), respectively. The solid and dashed lines denote $q$ and
$\phi$-propagators, respectively.}
\end{figure}

While Eqs.\ (\ref{eqs:4},\ref{eqs:5}) are sufficient to define our model, it is
conceptually helpful to consider the structure of a field theory that generates
these propagators. A Gaussian Landau-Ginzburg-Wilson (LGW) action consistent
with Gaussian propagators given by Eqs.\ (\ref{eqs:4}) has a structure
\bea
{\cal A}_0 &=& \int d{\bm x}\ \phi\,(t - a\,\partial_{\bm x}^2)\,\phi +
\sqrt{T}\gamma_1\int d{\bm x}\ \phi\,q
\nonumber\\
&& + \int d{\bm x}\ q\,(-\partial_{\bm x}^2/G + H\vert\Omega_n\vert)\,q.
\label{eq:6}
\eea
Here we consider the fields $\phi$ and $q$ functions of position ${\bm x}$ and
$\Omega_n$, and we use a schematic notation that leaves out everything not
needed for power counting, including sums over Matsubara frequencies and the
distinction between powers of $T$ and $\Omega_n$. The coefficients $t$, $a$,
$G$, and $H$ are the same as in Eqs.\ (\ref{eqs:4}), and $\gamma_1 = \sqrt{C_1
a/G}$. The renormalizations given by Eqs.\ (\ref{eqs:5}) are produced by a
non-Gaussian term with a structure
\be
{\cal A}_{\text{int}} = \sqrt{T}\gamma_2 \int d{\bm x}\ \phi\,q^2,
\label{eq:7}
\ee
with $\gamma_2 = \sqrt{GC_2/a}$. It gives rise to renormalizations of $a$ and
$H$ by means of the diagrams shown in Fig.\ \ref{fig:1} \cite{q4_footnote}.

The above model shares all qualitative features with the theory of disordered
itinerant Heisenberg quantum ferromagnets that was considered in Refs.\
\cite{Belitz_et_al_2001a,Belitz_et_al_2001b}. In particular, the integral
equations (\ref{eqs:5}) are slightly simplified versions of those derived and
solved in Refs.\ \onlinecite{Kirkpatrick_Belitz_1992b,Belitz_et_al_2001b}. The
prefactors in Eqs.\ (\ref{eqs:5}) have been chosen such that the one-loop terms
quantitatively agree with those obtained from the full integral equations.

Since the perturbation theory is given to all orders by the integral equations
(\ref{eqs:5}), one can determine the critical behavior by solving the latter.
This was done in Ref.\ \onlinecite{Belitz_et_al_2001b}. In what follows we
instead analyze the theory by standard RG techniques \cite{Wilson_Kogut_1974,
Ma_1976, Zinn-Justin_1996} order by order in a loop expansion. As we will see,
this leads to important insights which one misses by solving the integral
equations, and which transcend the specific ferromagnetic quantum phase
transition problem that inspired our model.

We define the scale dimensions of a length $L$ and frequency $\Omega$ or
temperature $T$ to be $[L]=-1$, $[\Omega]=[T]=z$, with $z$ a so far
undetermined dynamical critical exponent \cite{z_footnote}. The scale
dimensions of the fields $\phi$ and $q$ we define such that $a$ and $G$ are
dimensionless. We further define dimensionless coupling constants $h$, $c_1$,
and $c_2$ by
\be
H = \Lambda^{2-z}h\quad,\quad C_1 = \Lambda^{4-z}c_1\quad,\quad C_2 =
\Lambda^{2+\epsilon - z}c_2,
\label{eq:8}
\ee
and derive RG flow equations for these quantities.

The properties of our model allow us to determine several flow equations and
exponent relations exactly. Since $G$ is not renormalized, the field $q$ does
not carry a field renormalization, which means that the diffusive propagators
retains its Gaussian wave number dependence. $\phi$, however, does carry a
field renormalization, which gives rise to the critical exponent $\eta$ defined
by $a({\bm p}\to 0)\propto \vert{\bm p}\vert^{-\eta}$. The fact that $C_1$ and
$t$ are not renormalized implies the exact flow equations
\be
dc_1/d\ell = (4-z-\eta)c_1\quad,\quad dt/d\ell = (2-\eta)t.
\label{eq:9}
\ee
Both $z$ and the correlation length exponent $\nu$ (defined by $t\propto
b^{1/\nu}$) can therefore be expressed in terms of $\eta$ and are given by Eq.\
(\ref{eq:2b}).
The critical properties of the model are thus determined by only one
independent exponent, e.g., $\eta$. Power counting shows that $C_2$ is not
singularly renormalized either, so the flow equation for $c_2$ is also known
exactly,
\be
dc_2/d\ell = -(2-\epsilon)c_2.
\label{eq:11}
\ee
The flow equation for $h$, as well as $\eta$, need to be constructed in terms
of a loop expansion.

Let us first consider the theory to zero-loop order. Since $c_2$ is irrelevant
for all $d>2$, Eq.\ (\ref{eq:11}), there is a Gaussian FP with $\eta=0$, $z=4$.
This is the FP found by Hertz \cite{Hertz_1976}. It is not stable, as was first
shown in Ref.\ \onlinecite{Kirkpatrick_Belitz_1996}. In the present context,
this can be seen as follows.

Consider the perturbation theory as generated by iterating the integral
equations (\ref{eqs:5}). For our present purposes it suffices to do the
integrals in $d=4$. With $U=C_2/aH$, the perturbation theory then has the
structure
\bse
\label{eqs:12}
\bea
a({\bm k})/a &=& 1 + U F^{(1)}(\Lambda/\vert{\bm k}\vert) + U^2 {\tilde
F}^{(2)}(\Lambda/\vert{\bm k}\vert)
\nonumber\\
&&\hskip 90pt + O(U^3),
\label{eq:12a}\\
H(\Omega)/H &=& 1 + U\,G^{(1)}(\Lambda/\omega) + U^2\,G^{(2)}(\Lambda/\omega)
\nonumber\\
&&\hskip 90pt + O(U^3),
\label{eq:12b}
\eea
\ese
The proper expansion parameter is thus not $C_2$, but rather $U$, or its
dimensionless counterpart $u = c_2/ah$, which for $d<4$ is relevant with
respect to the Gaussian FP with a scale dimension of $\epsilon$, as can be seen
from Eq.\ (\ref{eq:8}). The Gaussian FP is therefore stable only for $d>4$,
despite the irrelevancy of $c_2$, since $h$ acts as a DIV \cite{DIV_footnote}.

To one-loop order, $F^{(1)}(x) = f_1 \ln x$ with $f_1 = 1/6$, and
$G^{(1)}(x\!\to\!\infty) = g_1 \ln x + O(x^{-4})$ with $g_1 = 3/4$. Via
standard methods \cite{methods_footnote} this yields Eqs.\
(\ref{eq:1},\ref{eqs:2}).

At two-loop order, ${\tilde F}^{(2)}(x\!\to\!\infty) = {\tilde f}_{2,2}\ln^2 x
+ O(\ln x)$ with ${\tilde f}_{2,2} = -3/32$, and $G^{(2)}(x\!\to\!\infty) =
g_{2,2}\ln^2 x + g_{2,1}\ln x + O(1)$ with $g_{2,2} = -1/16$, $g_{2,1} = 1/32$.
It is obvious, in either the momentum-shell or the field-theoretic RG methods,
that the coefficients of the two-loop $\ln^2$-terms must not be independent of
the coefficients of the one-loop $\ln$-terms if the theory is to be
renormalizable (in the latter method), or lead to scale-independent flow
equations and critical exponents (in either method). A straightforward analysis
shows that the condition is ${\tilde f}_{2,2} = g_{2,2} = -f_1 g_1/2$. This
condition is obeyed by $g_{2,2}$, but not by ${\tilde f}_{2,2}$. This problem
is germane to the structure of the perturbation theory and cannot be cured by
introducing additional renormalization constants (within the field-theoretic
method). To see its origin, let us analyze the contributions to the
$\ln^2$-terms.

Equation (\ref{eq:5b}) for $H(\Omega)$ in our model has the same structure as
insertions in a classical theory, since there is no frequency integral. The
one-loop correction to $a({\bm p})$ is linear in $\ln (\Lambda/p) = \ln\Lambda
- \ln p$ with $p=\vert{\bm p}\vert$. Both of these terms contribute to the
$\ln^2\Lambda$ contribution to $H(\Omega)$, with relative prefactors of $1$ and
$-1/2$, respectively. This structure guarantees that $g_{2,2} = -f_1g_1/2$. The
one-loop contribution to $H(\Omega)$, on the other hand, is linear in $\ln
(\Lambda/\Omega^{1/4})$, which leads to contributions to the $\ln^2\Lambda$
term in $a({\bm k})$ with relative prefactors of $1$ and $-1/4$, respectively.
As a result, one has ${\tilde f}_{2,2} = -(1-1/4)f_1g_1 = -3f_1g_1/4$. This is
a direct consequence of the existence of two time scales: If one replaced the
$\Omega$ in the integrand in Eq.\ (\ref{eq:5b}) by $\Omega^2$, which would
result in a model with only one time scale, $\Omega\sim {\bm p}^2$, one would
get ${\tilde f}_{2,2} = g_{2,2} = -f_1g_1/2$.

These observations suggest the following interpretation of this failure of the
renormalization procedure. After scaling the integration variable $\omega$ in
Eq.\ (\ref{eq:5a}) with $p^2/H$, the one-loop correction to $H(\omega)$ is
proportional to $\ln (\Lambda^4H/C_1\omega p^2)$. Writing this as
$\ln(\Lambda^4/p^4) - \ln\omega + \ln (Hp^2/C_1)$, we can write ${\tilde
F}^{(2)}$ in Eq.\ (\ref{eq:12a}) as
\be
{\tilde F}^{(2)}(\Lambda/\vert{\bm k}\vert) = F^{(2)}(\Lambda/\vert{\bm
k}\vert) - \frac{1}{32}\,\int_{\vert{\bm k}\vert}^{\Lambda}dp\,\frac{1}{p}\,\ln
w(p),
\label{eq:13}
\ee
where $F^{(2)}(x\to\infty) = -(\ln x)^2/16 + O(\ln x)$, and $w(p) = Hp^2/C_1 =
(h/c_1)(p/\Lambda)^2$. This leads to the two-loop flow equations (\ref{eqs:3}),
where $w\! =\! w(p\! =\! \Lambda)\! =\! h/c_1$. The FP value of the irrelevant
variable $w$ is $w^*=0$. Inserted into the $u$-flow equation, this yields an
infinite coefficient at two-loop order. We also see that the physical origin of
this problem is the second (diffusive) time scale, which is reflected in the
scale dimension of $w\propto h$.

The following picture now emerges. $h$ is a DIV since the loop expansion is in
powers of $u = c_2/ah$. At one-loop order, this makes the Gaussian FP unstable,
but one has the seemingly sensible FP discussed in connection with Eqs.\
(\ref{eq:1},\ref{eqs:2}). The two-loop term, however, goes as $h^{-2}\ln h$
rather than just as $1/h^2$, and this leads to a breakdown of the loop
expansion. If one eliminates $w$ by solving Eq.\ (\ref{eq:3b}) and substituting
the result in Eq.\ (\ref{eq:3a}), one obtains scale dependent flow equations
and critical exponents. As we have seen, this is a consequence of the presence
of two time scales in conjunction with the structure of the perturbation theory
for our model. It is therefore not possible to analyze this field theory by
means of standard RG methods. At higher order in the loop expansion, higher
powers of $\ln w$ will appear. Clearly, a resummation to all orders is needed
in order to determine the critical behavior. Indeed, a solution of the integral
equations, which achieve such a resummation, shows that the asymptotic critical
behavior consists of power laws multiplied by log-normal terms
\cite{Belitz_et_al_2001b}.

In summary, we have used the quantum ferromagnetic transition in disordered
metals to show that one cannot necessarily trust low-order perturbative RG
results, even in the absence of any indications that the perturbative FP one
has found might not be stable. The mechanism we have discussed involves a DIV
and is rather general, and is thus expected to be operative at other phase
transitions as well. For instance, our analysis casts doubt on a recent
one-loop calculation for the clean quantum ferromagnetic problem
\cite{Kirkpatrick_Belitz_2003}. We note, however, that the presence of multiple
time scales, while necessary, is not sufficient to produce the effect. For
instance, at the liquid-gas critical point mentioned in the introduction, the
analog of the irrelevant variable $w$ is not dangerous
\cite{DeDominicis_Peliti_1978}.

This work was supported by the NSF under grant Nos. DMR-01-32555 and
DMR-01-32726, and by a fellowship of the Deutsche Forschungsgemeinschaft
(J.R.). Part of this work was performed at the Aspen Center for Physics.

\vskip -0mm

\begin{thebibliography}{22}
\bibitem[{Her()}]{Hertz_footnote}
\bibinfo{note}{According to a tree-level analysis, $z=4$ implies an upper
  critical dimension $d_{\text c}^+ = 0$ \cite{Hertz_1976}. This is fallacious,
  however, as can be seen from the fact that $y$ becomes relevant with respect
  to the mean-field fixed point for $d<4$. This is the first manifestation of
  the dangerous irrelevancy of $w$, as we will see below.}
\bibitem[{DIV()}]{DIV_footnote}
\bibinfo{note}{We use the term ``dangerous irrelevant variable'' in a very
  general sense: The loop expansion parameter itself depends singularly on the
  irrelevant variable $h$, not just some scaling function.}
\bibitem[{met()}]{methods_footnote}
\bibinfo{note}{To one-loop order one can use either a momentum-shell RG, or a
  field-theoretic one, to derive these results. At two-loop order, only the
  field-theoretic method is practical, as the shell integrals become too
  complicated.}
\end{thebibliography}

\end{document}